\documentclass[
    aps,
    prl,
    twocolumn,
    superscriptaddress,
    showpacs,
    floatfix,
    longbibliography 
]{revtex4-2}

\usepackage{amsmath,amssymb,amsfonts} 
\usepackage{graphicx}                 
\usepackage{dcolumn}                  
\usepackage{bm}                       
\usepackage{hyperref}                 
\usepackage{color}                    
\usepackage[utf8]{inputenc}
\usepackage[T1]{fontenc}
\usepackage{float}
\usepackage{placeins} 

\usepackage{balance}

\hypersetup{
    colorlinks=true,
    linkcolor=blue,
    filecolor=magenta,      
    urlcolor=blue,
    citecolor=blue,
}

\begin{document}

\title{Dynamical Partition Functions of Stochastic Dynamics via Variational Flows}

\author{Zequn Lin}
\email{linzequn@westlake.edu.cn}
\affiliation{Department of Physics, Fudan University, Shanghai 200433, China}
\affiliation{Center for Interdisciplinary Studies, Westlake University, Hangzhou 310030, China}

\author{Ying Tang}
\email{jamestang23@gmail.com}
\affiliation{Institute of Fundamental and Frontier Sciences, University of Electronic Science and Technology of China, Chengdu 611731, China}
\affiliation{Non-classical Information Science Basic Discipline Research Center of Sichuan Province, University of Electronic Science and Technology of China, Chengdu 611731, China}

\begin{abstract}
Nonequilibrium thermodynamics is governed by the dynamical partition function, and its computation in high-dimensional continuous-state dynamics is a longstanding challenge. The Feynman-Kac formalism provides a rigorous representation for generating functions of arbitrary path observables; however, practical evaluation beyond low dimensions or the weak-noise limit is hindered by the curse of dimensionality and the exponentially growing replica demands of trajectory-based methods. Here we develop a mesh-free neural variational framework that realizes the Feynman-Kac theorem with generative flow models, recasting tilted stochastic evolution as a time-dependent optimization problem. This approach enables the direct computation of both finite-time and asymptotic trajectory thermodynamics in a unified manner. The method applies to general observables, enabling the evaluation of work, entropy production, and current fluctuations. We demonstrate the accuracy and scalability of this method in various nonequilibrium systems including high-dimensional cases. 
\end{abstract}

\maketitle

\textit{Introduction---} Continuous-state stochastic dynamics \cite{duan2015introduction} provides a fundamental description for systems ranging from molecular motor transport \cite{julicher1997modeling, zhang2012stochastic}, active matter \cite{bechinger2016active,romanczuk2012active,bonilla2019active,tailleur2008statistical}, stochastic thermodynamics \cite{seifert2008stochastic,seifert2012stochastic}, and machine learning \cite{feng2021inverse,yang2023stochastic,zhang2025heavy,ikeda2025speed}. Their evolution is typically described by stochastic differential equations (SDEs) or Fokker-Planck equations (FPEs). To fully characterize these dynamics, it is essential to use trajectory ensembles, particularly when these systems violate detailed balance and sustain probability flux \cite{ao2004potential, wang2008potential, tang2018escape}. Central to this trajectory-based framework is the dynamical partition function, $Z_t(\lambda)$. Analogously to the canonical partition function in equilibrium, $Z_t(\lambda)$ serves as a universal generating functional. Coupling the dynamics with specific path observables—such as thermodynamic work in driven systems \cite{seifert2012stochastic, chen2025coercivity}, entropy production quantifying irreversibility \cite{crooks1999entropy, wang2016entropy,PhysRevE.101.022129, aguilera2026inferring}, and winding numbers representing particle currents \cite{ferre2018adaptive}—allows one to derive a broad class of thermodynamic quantities, ranging from finite-time statistics \cite{jarzynski1997nonequilibrium, crooks1999entropy} to asymptotic large deviations \cite{touchette2009large,ge2012analytical}.

Despite its fundamental importance, the exact solutions for $Z_t(\lambda)$ are generally intractable and require numerical estimations \cite{risken1996fokker}. Standard mesh-based approaches for solving FPE, such as finite element or spectral methods, suffer severely from the curse of dimensionality in high-dimensional systems. Traditional trajectory simulation methods, e.g., cloning algorithms \cite{2006Direct, grassberger2002go, lecomte2007numerical}, are plagued by an exponentially large replica size when sampling rare fluctuations. Meanwhile, recent advances in the intersection of machine learning and stochastic dynamics \cite{yan2021learning, reh2022variational, Boffi_2023, tang2024learning,tang2023neural,klinger2025computing} offer a promising strategy for this problem, particularly a type of generative model called normalizing flows (NF) \cite{dinh2016density, bekri2025flowkac, feng2021solving, reh2022variational}. These methods successfully capture the evolution of probability distributions $p(\mathbf{x}, t)$ for a given FPE. However, because they are designed primarily to approximate normalized probability distributions, they are not directly applicable to calculating the dynamical partition function and related thermodynamic observables.

In this Letter, we present a variational neural framework designed to compute the dynamical partition function in a mesh-free manner. By synthesizing the time-dependent variational principle \cite{PhysRevLett.127.230501} (TDVP) for normalizing flows \cite{reh2022variational} with the rigorous Feynman-Kac formalism \cite{hummer2001free, chetrite2015nonequilibrium}, we transform the estimation of $Z_t(\lambda)$ from a passive integration task into an intrinsic optimization problem. To the best of our knowledge, this is the first implementation of a neural dynamical evolution scheme capable of calculating the dynamical partition function for general continuous-state stochastic systems. We demonstrate the applicability and efficiency of this framework across a set of nonequilibrium systems. We first calculate the  work in a two-dimensional Ornstein-Uhlenbeck process. This result verifies the Jarzynski equality in a system that violates detailed balance. Moving beyond analytically solvable models, we compute large deviations in a periodic diffusion system \cite{ferre2018adaptive}. Finally, we compute entropy production in a 16-dimensional system with random orthogonal couplings. This system serves as a high-dimensional benchmark lacking analytical solutions \cite{wu2025computing}.

\textit{Stochastic dynamics and path observables---} We consider a continuous-state stochastic system modeled by the SDE: 
\begin{equation}
d\mathbf{x}_t = \mathbf{f}(\mathbf{x}_t)\,dt + \sqrt{2\mathbf{D}} \, d\mathbf{W}_t,
\label{eq:sde}
\end{equation} where $\mathbf{x}_t \in  \mathbb{R}^d$ denotes the state vector, $\mathbf{D}$ is the diffusion matrix and $\mathbf{W}_t$ is the standard Brownian motion. To characterize the fluctuating thermodynamics of such systems, we focus on a generalized time-integrated path observable defined via Stratonovich integration ($\circ$):
\begin{equation}
O_t = \int_0^t f(\mathbf{x}_\tau) d\tau + \int_0^t \mathbf{g}(\mathbf{x}_\tau) \circ d\mathbf{x}_\tau.
\label{eq:observable}
\end{equation}This formulation is ubiquitous in stochastic thermodynamics, encompassing purely state-dependent quantities (e.g., thermodynamic work) as well as increment-dependent quantities (e.g., entropy production and particle currents). The full counting statistics of this dynamical observable are encoded in the dynamical partition function $Z_t(\lambda) = \langle e^{-\lambda O_t} \rangle$, with the expectation taken over all possible stochastic trajectories.

\begin{figure*}[htbp]
    \centering
    \includegraphics[width=\textwidth]{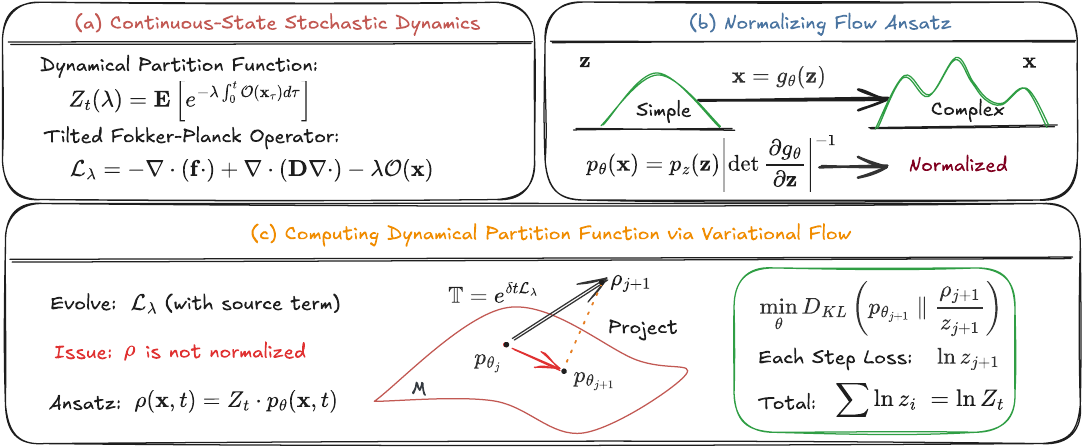}
    \caption{Computational framework for evaluating the dynamical partition function via normalizing flow. (a) Continuous-state stochastic dynamics are governed by a tilted Fokker-Planck operator $\mathcal{L}_\lambda$, which introduces a source term $-\lambda\mathcal{O}(\mathbf{x})$ and breaks probability conservation. The central goal is to compute the dynamical partition function $Z_t(\lambda)$. (b) The normalizing flow ansatz parameterizes a  normalized probability distribution $p_\theta(\mathbf{x})$ by applying an invertible transformation to a simple latent distribution $p_z(\mathbf{z})$, providing exact likelihood computation via the Jacobian determinant. (c) The decoupled variational training loop on the probability manifold $\mathcal{M}$. The unnormalized density $\rho$ is decomposed into a global scaling factor $Z_t$ and a normalized shape $p_\theta$. In each discrete time step, the short-time transition operator $\mathbb{T} = e^{\delta t \mathcal{L}_\lambda}$ evolves the state off the manifold to an unnormalized target $\rho_{j+1}$. The network parameters are updated by projecting this target back onto $\mathcal{M}$, minimizing the Kullback-Leibler divergence against the explicitly normalized state $\rho_{j+1}/z_{j+1}$. The residual of this optimization directly yields the single-step partition function increment $\ln z_{j+1}$, which is accumulated over time to evaluate the total $Z_t$.}
    \label{protocol}
\end{figure*}

\textit{Generalized Feynman-Kac formalism---} Evaluating $Z_t(\lambda)$ through direct trajectory sampling suffers from exponentially growing variance for rare fluctuations. Instead, the Feynman-Kac formula maps this path-integral expectation to a partial differential equation (PDE). To handle generalized observables containing stochastic increments ($\mathbf{g} \neq 0$, e.g., entropy production), we employ the Cameron-Martin-Girsanov transformation \cite{chetrite2015nonequilibrium} to absorb these path fluctuations into an effective drift. This universal procedure ensures that any path observable can be mapped to a standard time-integrated form. Consequently, $Z_t(\lambda)$ can be obtained as the spatial integral of an unnormalized density, $Z_t(\lambda) = \int \rho(\mathbf{x}, t) d\mathbf{x}$, where $\rho(\mathbf{x}, t)$ evolves under the tilted FPE, $\partial_t \rho = \mathcal{L}_\lambda \rho$ (FIG.~\ref{protocol}(a)). The full generator $\mathcal{L}_\lambda$ encapsulates these Girsanov corrections (derived in the supplemental material). For the specific subclass of purely state-dependent observables ($\mathbf{g}=0$, $f=\mathcal{O}$), the tilted operator  simplifies to the standard form:
\begin{equation}
\mathcal{L}_\lambda \rho = -\nabla \cdot (\mathbf{f}\rho) + \nabla \cdot (\mathbf{D}\nabla\rho) - \lambda \mathcal{O}(\mathbf{x})\rho.
\label{eq:tilted}
\end{equation}

\textit{Variational neural ansatz---} Solving the tilted FPE exactly is generally intractable in high dimensions. To enable a mesh-free variational inference, we decompose the unnormalized density into a scalar weight and a  normalized probability density: $\rho(\mathbf{x}, t) = Z_t(\lambda) p(\mathbf{x}, t)$. We then parameterize $p(\mathbf{x},t)$ using NF as an ansatz $p_\theta(\mathbf{x})$. By mapping a tractable reference distribution through a sequence of invertible neural network layers $\mathbf{x} = g_\theta(\mathbf{z})$, NF provides exact likelihood evaluations via the change-of-variables formula (FIG.~\ref{protocol}(b)). This formulation avoids high-dimensional spatial integrals.

\textit{Calculate dynamical partition functions---} To evolve the neural ansatz, we formulate the progression of the tilted FPE as a continuous-time variational inference problem. The framework minimizes the Kullback-Leibler (KL) divergence between the parameterized manifold and the exact evolved state. Formally, we define the short-time transition operator $\mathbb{T} = e^{\delta t \mathcal{L}_\lambda}$. The exact unnormalized target at step $j+1$ is $\rho_{j+1} = \mathbb{T} p_{\theta_j}$. We project this back onto the manifold by minimizing $D_{KL} \left( p_{\theta_{j+1}} \parallel \frac{\rho_{j+1}}{z_{j+1}} \right)$. While various optimization schemes can be employed, executing this minimization in the continuous-time limit projects the exact physical dynamics onto the tangent space of the neural manifold. This projection translates the tilted FPE into a natural gradient descent over the parameters $\theta$:
\begin{equation}
\label{eq:tdvp_core}
\dot{\theta} = \mathbf{F}_\theta^{-1} \mathbb{E}_{\mathbf{x} \sim p_\theta} \left[ (\nabla_\theta \ln p_\theta) E_{\text{loc}}(\mathbf{x}) \right],
\end{equation}
where $\mathbf{F}_\theta$ is the Fisher information matrix acting as the local metric, and $E_{\text{loc}} \equiv p_\theta^{-1}\mathcal{L}_\lambda p_\theta$ is the instantaneous local energy dictating the spatial deformation (FIG.~\ref{protocol}(c)). The scalar residual of this KL minimization  determines the growth rate of the partition function: $\partial_t \ln Z_t = \mathbb{E}_{p_\theta}[E_{\text{loc}}]$. Accumulating this scalar rate over time alongside the tilted operator evolution directly yields the exact finite-time dynamical partition function $Z_t(\lambda)$. Furthermore, once the optimization reaches a steady state, we extract the asymptotic scaled cumulant generating function (SCGF) $\psi(\lambda) = \lim_{t \to \infty} t^{-1} \ln Z_t(\lambda)$. 

\textit{Applications---} We demonstrate the capability and scalability of our neural framework across three distinct classes of nonequilibrium systems. The dynamics of all three can be unified under the general overdamped Langevin dynamics:
\begin{equation}
    d\mathbf{x}_t = \left[ -\nabla V(\mathbf{x}_t, t) + \mathbf{b}(\mathbf{x}_t, t) \right] dt + \sqrt{2\mathbf{D}} \, d\mathbf{W}_t.
    \label{eq:general_sde}
\end{equation}
Here, $V$ represents the confining potential, and $\mathbf{b}$ denotes the non-conservative drift field, which breaks detailed balance.

\textit{Finite-time Jarzynski equality ($d=2$)---} To benchmark our framework against analytical results in nonequilibrium thermodynamics, we first investigate a coupled linear system driven by a time-dependent protocol. We map this system to Eq.~(\ref{eq:general_sde}) by setting $\mathbf{D}=\mathbf{I}$ and decomposing the linear drift. The potential is harmonic, $V(\mathbf{x}) = \frac{1}{2}\|\mathbf{x}\|^2$ (with $\|\cdot\|$ the Euclidean norm), while the non-conservative drift introduces rotational coupling and external driving:
\begin{equation}
    \mathbf{b}(\mathbf{x}, t) = \mathbf{\Omega} \mathbf{x} + \boldsymbol{\mu}(t), \quad 
    \mathbf{\Omega} = \begin{pmatrix} 0 & 1 \\ -1 & 0 \end{pmatrix}.
\end{equation}
The skew-symmetric matrix $\mathbf{\Omega}$ induces a probability current. The external protocol is defined as $\boldsymbol{\mu}(t) = [\mu_1(t), 0]^\top$, where $\mu_1(t) = 20t$ for $t \le 0.025$ and $\mu_1(t) = 0.5$ thereafter. This protocol drives energy transduction across the degrees of freedom.

\begin{figure}[htbp]
    \centering
    \includegraphics[width=\linewidth]{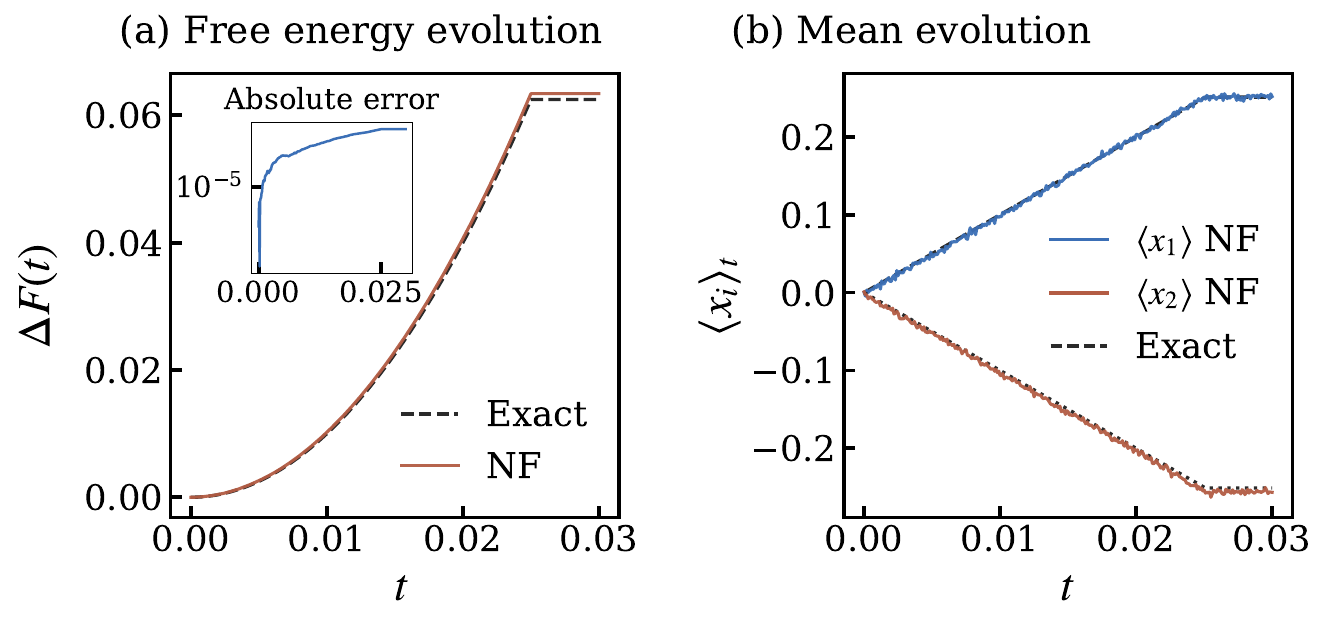}
    \caption{Time evolution of the two-dimensional OU process driven by a protocol force. (a) The red curve represents the free energy difference $\Delta F(t)$ estimated from normalizing flow. The dashed black curve denotes the exact solution.  Inset: Absolute error. (b) Trajectories of the first moments $\langle x_1 \rangle_t$ and $\langle x_2 \rangle_t$ of the biased ensemble under drift protocol $\boldsymbol{\mu}(t)$. Solid lines (blue and red) indicate the empirical means estimated from flow model at each time step. Black dashed and dotted lines correspond to the exact evolution. }
    \label{OU}
\end{figure}

To evaluate the excess work $W$ performed by the external protocol on this driven nonequilibrium process, we configure the path observable as the injected power associated with the effective potential, $\mathcal{O}(\mathbf{x}_t) = \mathbf{x}_t^\top\mathbf{U}^\top \mathbf{F}^{-1} \dot{\boldsymbol{\mu}}(t)$ \cite{hatano2001steady}. When the tilting parameter is set to $\lambda=1$, the dynamical partition function directly yields the Jarzynski equality \cite{jarzynski1997nonequilibrium}, $\langle e^{-W} \rangle = e^{-\Delta F}$, where $\Delta F$ is the free energy difference. The estimate of the instantaneous free energy $\Delta F(t) = -\ln Z_t(1)$ matches the exact solution (supplemental material) throughout the driving protocol ($K=20$). The absolute error remains below $10^{-4}$ (FIG.~\ref{OU}(a)).  This precision comes without the exponential variance explosion of trajectory cloning. Because the normalizing flow maintains a normalized distribution, it provides continuous access to the biased ensemble. Figure~\ref{OU}(b) illustrates the evolution of the first moments, $\langle x_1 \rangle_t$ and $\langle x_2 \rangle_t$. Although the external protocol $\boldsymbol{\mu}(t)$ solely drives the $x_1$ coordinate, the skew-symmetric coupling $\mathbf{\Omega}$ induces a transverse response in $x_2$. The neural ansatz captures the rotational coupling, consistent with the exact evolution of the first moments.

\textit{Topological current ($d=1$)---} We next verify the solver's capability on non-Euclidean manifolds by simulating diffusion on the unit circle \cite{ferre2018adaptive}. The state dynamics follow Eq.~(\ref{eq:general_sde}) with scalar diffusion $D=1$. The conservative force is derived from a periodic potential $V(x) = \cos(2\pi x)$, while the non-conservative drift is a constant drive along the manifold:
\begin{equation}
    b(x) = \gamma, \quad \gamma \in \mathbb{R}.
\end{equation} In our numerical implementation, we set $\gamma = 1$.

\begin{figure}[h]
    \centering
    \includegraphics[width=\linewidth]{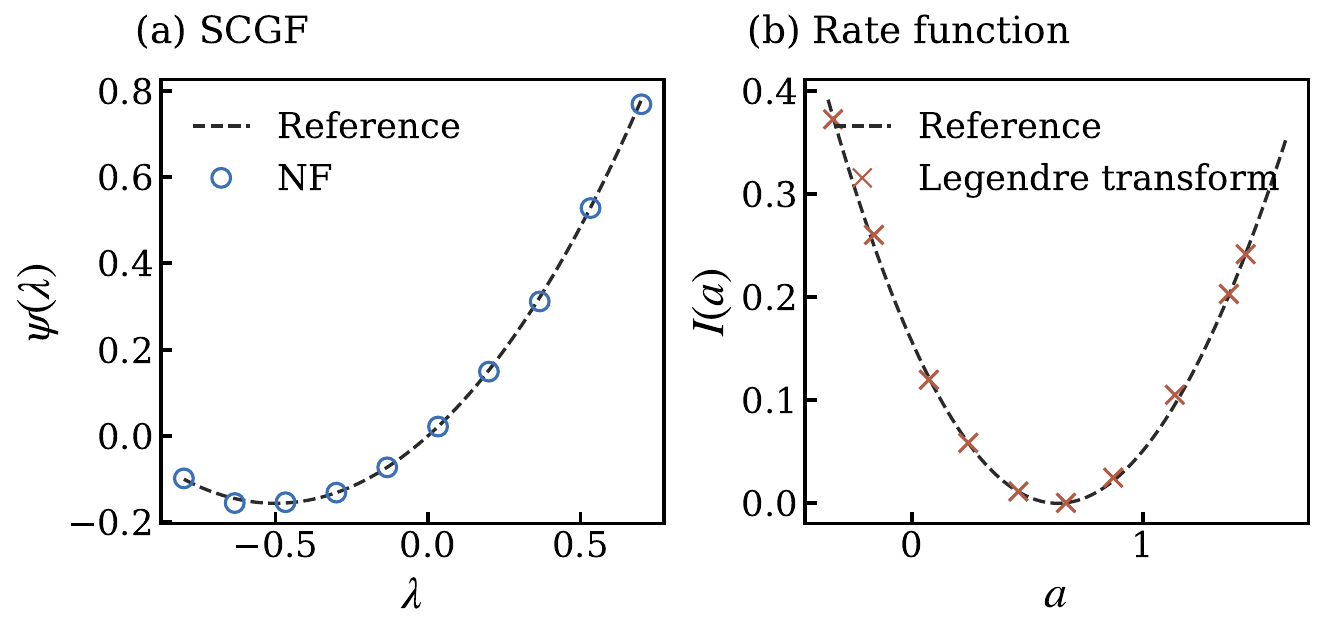}
    \caption{Large deviations for a nonequilibrium driven diffusion process on the unit circle $\mathbb{S}^1$. (a) The SCGF $\psi(\lambda)$ of the time-averaged current versus the conjugate tilting parameter $\lambda$. Blue open circles ($\circ$) indicate discrete evaluations obtained from our framework (computed at 10 points). The black dashed line is the reference computed via spectral diagonalization of the tilted generator. The  asymmetry of $\psi(\lambda)$ reflects the broken detailed balance induced by the constant non-conservative drive $\gamma$. (b) The corresponding rate function $I(a)$ governing the empirical current fluctuations $a$. Red crosses ($\times$) are derived from the numerical Legendre transform and match the reference (black dashed line). }
    \label{Hugo}
\end{figure}

 We evaluate the time-averaged winding current,
\begin{equation}
A_t = \frac{1}{t} \int_0^t 1 \circ dx_\tau,
\end{equation}
which quantifies the mean particle velocity on the periodic manifold, as an increment-dependent observable ($\mathbf{g} \neq 0$). Following the convention for current-type observables \cite{chetrite2015nonequilibrium, ferre2018adaptive}, we define the SCGF $\psi(\lambda) = \lim t^{-1} \ln \langle e^{\lambda tA_t} \rangle$, equivalent to $Z_t(-\lambda)$.  The exact forms of both the SCGF and the corresponding rate function remain unknown for this observable. By evolving the neural ansatz to steady state, we extract $\psi(\lambda)$ directly from the asymptotic growth rate of the dynamical partition function. As shown in FIG.~\ref{Hugo}(a), our neural SCGF quantitatively reproduces the curve computed via numerical spectral diagonalization. While diagonalization provides a rigorous baseline here, its exponential scaling makes it intractable for high-dimensional manifolds \cite{ferre2018adaptive}. Applying the Legendre transform, $I(a) = \sup_\lambda \bigl[ \lambda a - \psi(\lambda) \bigr]$, we obtain the large deviation rate function (FIG.~\ref{Hugo}(b)). This agreement validates the ability of our framework to calculate large deviations in periodic topologies.

\textit{High-dimensional entropy production ($d=16$)---} To demonstrate scalability, we consider a 16-dimensional overdamped Langevin process ($\mathbf{D}=\epsilon\mathbf{I}$) where a quartic confining potential competes with a divergence-free rotational drift \cite{wu2025computing}:
\begin{align}
V(\mathbf{x}) &= \frac{1}{2}\mathbf{x}^\top \mathbf{M} \mathbf{x} + 4\|\mathbf{x}\|^4, \label{eq:potential}\\
\mathbf{b}(\mathbf{x}) &= \eta(\|\mathbf{x}\|) \mathbf{B} \mathbf{x}.
\label{eq:drift}
\end{align}
Here, $\mathbf{M} = \text{diag}(5, 6, \dots, 20)$, and the skew-symmetric matrix $\mathbf{B} = \mathbf{Q}^\top \text{diag}(\mathbf{B}_1, \dots, \mathbf{B}_8) \mathbf{Q}$ comprises $2\times 2$ blocks $\mathbf{B}_k = [\begin{smallmatrix} 0 & 1 \\ -1 & 0 \end{smallmatrix}]$ rotated by a random orthogonal matrix $\mathbf{Q}$. A smooth cutoff $\eta(r)$ (unity for $r \le 1$, zero for $r \ge 2$) ensures global stability. This interplay between the anisotropic confinement and the high-dimensional rotational drift generates a complex, non-gradient steady-state probability current. We quantify macroscopic irreversibility via the steady-state entropy production, $O_t = \frac{1}{\epsilon} \int_0^t \mathbf{b}(\mathbf{x}_\tau) \circ d\mathbf{x}_\tau$, and compute its SCGF, $\psi(\lambda) = \lim_{t \to \infty} \frac{1}{t} \ln \langle e^{-\lambda O_t} \rangle$. No exact expression exists for either the SCGF or the rate function. We operate at $\epsilon = 0.1$. 

\begin{figure}[htbp]
    \centering
    \includegraphics[width=\linewidth]{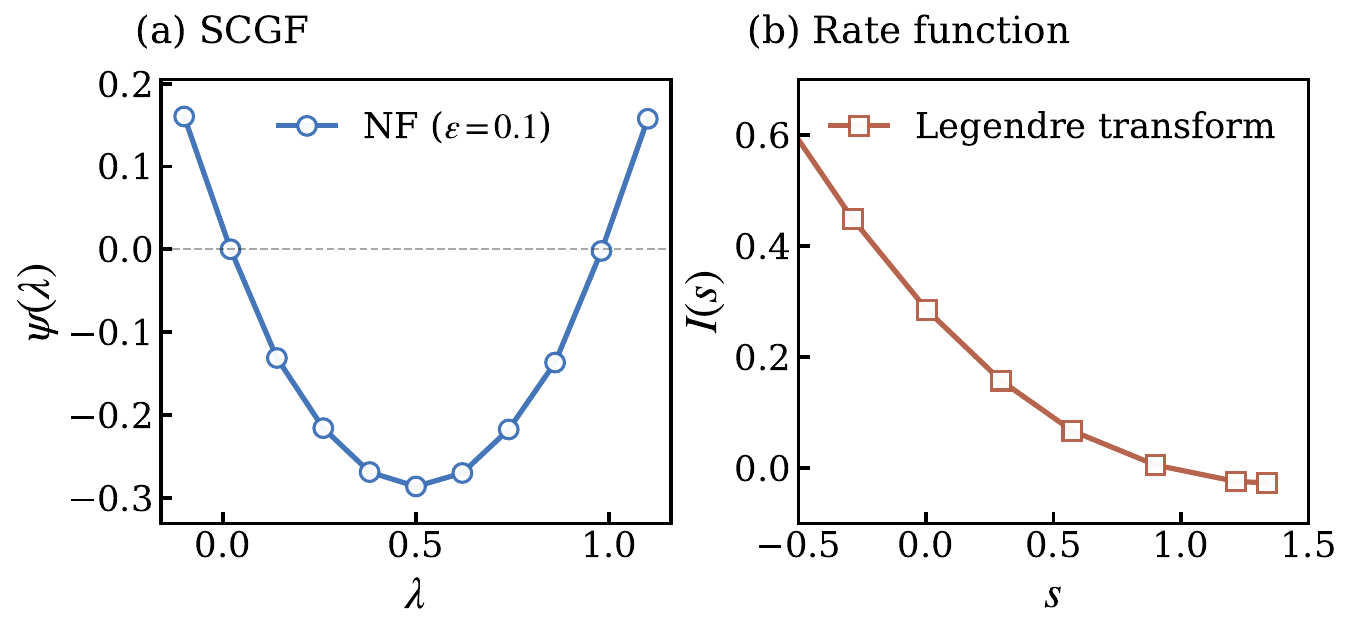}
    \caption{Large deviations of entropy production for the 16-dimensional diffusion process at $\epsilon = 0.1$. (a) The SCGF $\psi(\lambda)$ computed via the neural ansatz (blue open circles ($\circ$)) quantitatively matches the benchmark results  from Fig.~7 in Ref.~\cite{wu2025computing}. (b) The corresponding rate function $I(s)$ obtained via numerical Legendre transform.}
    \label{HK}
\end{figure}

 The computed SCGF (FIG.~\ref{HK}(a)) matches the benchmark of Fig.~7 in Ref. \cite{wu2025computing}. The Legendre-transformed rate function $I(s)$ (FIG.~\ref{HK}(b)) agrees with reference results. Together with the 100-dimensional benchmark \cite{Boffi_2023} (supplemental material), these results demonstrate scalability to high dimensions.

\textit{Global-in-time variational framework---}
Besides the sequential TDVP projection Eq.~(\ref{eq:tdvp_core}), one can implement the variation on the full time region~\cite{wang2026quantum}. This variation may leverage a parameterized spatiotemporal distribution $p_\theta(\mathbf{x},t)$, treating time $t$ as an input to the normalizing flow. On a discrete grid $t_j=j\delta t$, this global ansatz yields a time-averaged Kullback-Leibler divergence:
\begin{equation}
\overline{\mathcal{D}}[\theta]=\frac{1}{T}\sum_{j=0}^{N-1}
D_{KL}\!\left[p_\theta(\cdot,t_{j+1})\,\Big\|\,
\frac{\mathbb{T}_j p_\theta(\cdot,t_j)}{z_{j+1}}\right].
\end{equation}
Because the flow model is normalized, this gives a variational lower bound for the dynamical partition function:
\begin{equation}
\frac{1}{T}\ln Z_T = \overline{\mathcal{D}}[\theta] - \frac{1}{T}\sum_{j=0}^{N-1}L_{j+1} \ge -\frac{1}{T}\sum_{j=0}^{N-1}L_{j+1}, 
\label{global}
\end{equation}
where the local loss is defined as:
\begin{equation}
L_{j+1} \equiv \mathbb{E}_{\mathbf{x}\sim p_\theta(\cdot,\,t_{j+1})}
\left[ \ln p_\theta(\mathbf{x},t_{j+1}) - \ln\big(\mathbb{T}_j p_\theta(\mathbf{x},t_j)\big) \right].
\end{equation}
Maximizing this bound provides a direct route to evaluate $\ln Z_T$ globally, which may have a tradeoff on the accuracy and efficiency compared with the sequential TDVP.

\textit{Discussion---} In summary, we have presented a variational neural framework to compute the dynamical partition function and large deviations for continuous-state stochastic systems. Using normalizing flows and the Feynman-Kac formalism, we recast tilted stochastic evolution as an optimization problem. This mesh-free method effectively avoids the exponentially growing replica demands, which plague trajectory cloning methods. In the supplemental material, we validate this computational scalability. Specifically, we track the unbiased dynamics ($\lambda=0$) of a 100-dimensional system to confirm the architecture's high-dimensional capacity. Complexity analysis and empirical scaling tests confirm that the TDVP projection scales polynomially as $\mathcal{O}(d^2)$. This scaling was verified for single-step throughput up to $d=10^4$  on a single A100 GPU (end matter).

Our framework builds upon recent work \cite{reh2022variational} that successfully combined the TDVP and normalizing flows to solve the FPE. We extend this approach from probability-conserving dynamics to the explicit evaluation of the dynamical partition function via the Feynman-Kac formula. This yields a versatile framework for continuous-state stochastic systems. Recent Feynman-Kac solvers \cite{bekri2025flowkac} target probability-conserving FPEs through path-dependent regression, but this regression does not directly yield dynamical partition functions, and their trajectory dependence complicates rare-event sampling. Variational optimal control methods \cite{yan2021learning} learn effective forces, but their reliance on explicit trajectory integration makes finite-time observables expensive to extract.  These frameworks are primarily formulated for asymptotic ($t \to \infty$) large deviations and thus require extensively long simulated paths.

While normalizing flows offer exact likelihood evaluations, their strict bijectivity inherently constrains their representational capacity. One might overcome this limitation by replacing flows with unconstrained architectures, such as transformers \cite{boffi2024deep}, flow matching \cite{lipman2023flow}, or diffusion models \cite{song2021scorebased}. Sampling could then be augmented with molecular dynamics (MD) simulations or local Markov chain Monte Carlo (MCMC) methods \cite{asghar2024efficient, noe2019boltzmann, gabrie2022adaptive}. The global bound (Eq.~(\ref{global})) eliminates sequential time-stepping at the cost of higher memory; sequential variation is adopted here for lower memory.

Beyond analytical generators, extending our framework to data-driven inference represents a biophysical frontier. Recent breakthroughs reconstruct high-dimensional vector fields with growth \cite{zhang2025inferring} and non-conservative dynamics via unbalanced optimal transport \cite{zhang2025learning} directly from single-cell snapshots. Applying our variational framework to these inferred FPEs allows us to directly extract macroscopic thermodynamics.

\textit{Acknowledgments---} We thank Prof.~Hong Qian and Ms.~Miao~Chen for helpful discussions. This work is supported by the National Natural Science Foundation of China (Projects 12322501 and 12575035) and the Natural Science Foundation of Sichuan Province (Grant 2026NSFSCZY0124).

\textit{Code availability---} 
The PyTorch implementation of the proposed method will be available at
\url{https://github.com/linzequn17/NFDPF}.

%

\clearpage 
\onecolumngrid 

\begin{center}
\textbf{END MATTER}
\end{center}
\vspace{0.3cm}
\twocolumngrid
\textit{Computational Complexity and Empirical Scaling.---} 

\begin{figure}[H]
    \centering
    \includegraphics[width=\linewidth]{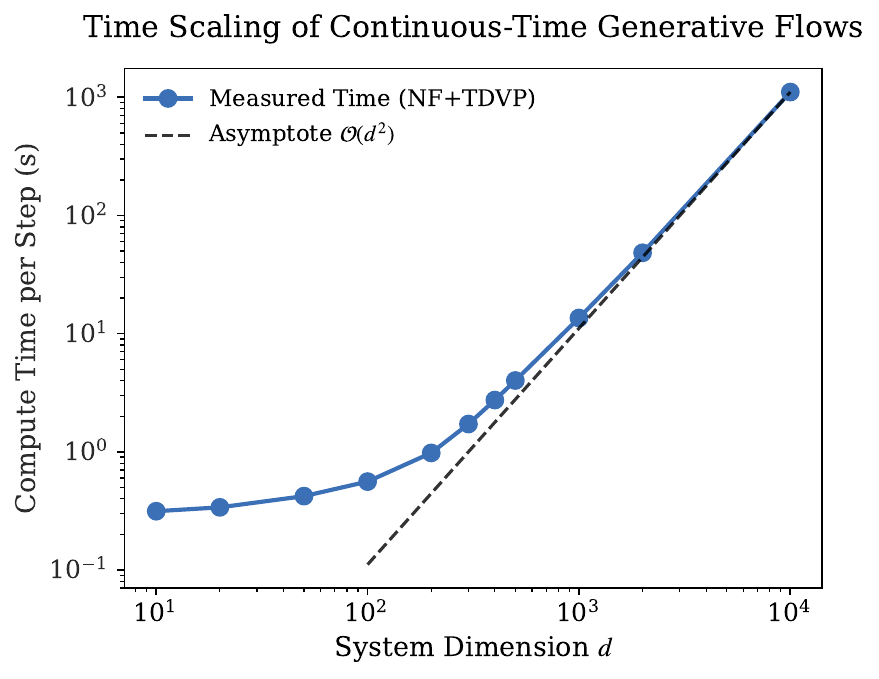}
     \caption{Scaling behavior of the computational cost with respect to system dimensionality $d$. The blue solid line shows the empirical wall-clock time per full evolution step, measured on a single NVIDIA A100 GPU with a fixed ensemble size of 8,000. The black dashed line indicates the theoretical $\mathcal{O}(d^2)$ polynomial scaling. The empirical performance aligns closely with this theoretical bound. }
    \label{fig:scaling}
\end{figure}

We analyzed the computational complexity of our neural variational framework to establish its scalability. Each time step consists of two primary phases. First, we evaluate the instantaneous local energy. This requires the  Laplacian of the log-density. Second, we solve the TDVP projection via the conjugate gradient (CG) method. The Laplacian evaluation, performed via automatic differentiation or centered finite differences, requires 2d independent forward passes through the normalizing flow. For a network with constant hidden layer capacity, a single pass scales as $\mathcal{O}(d)$. The Laplacian evaluation scales as $\mathcal{O}(d^2)$. The TDVP projection, which utilizes implicit Hessian-Vector Products, scales linearly as $\mathcal{O}(d)$. Therefore, in the high-dimensional limit, the overall asymptotic time complexity per evolution step is bounded by $\mathcal{O}(d^2)$.

For comparison, trajectory cloning algorithms estimate $\langle e^{\lambda t A_t}\rangle$ by simulating $N_c$ independent copies of the SDE, each for physical time $T$. The number of clones required to maintain a fixed relative error grows as $N_c \propto e^{\psi(\lambda)T}$ due to the exponential variance of the Feynman-Kac estimator. For extensive observables in $d$ dimensions, $\psi(\lambda) \propto d$, so $N_c \propto e^{c d}$. Multiplying by the $\mathcal{O}(d)$ per-trajectory integration cost, the total computational time of cloning scales as $\mathcal{O}(d e^{c d})$, in contrast to the polynomial $\mathcal{O}(d^2)$ scaling of our variational approach. This exponential barrier is a fundamental statistical bottleneck. Rare fluctuations are exponentially suppressed in high dimensions, and cloning must sample them by brute force. 

To empirically validate this polynomial bound, we benchmarked the actual wall-clock time for a single variational evolution step by scaling up a physical model: a nonequilibrium system of $N$ globally coupled two-dimensional particles driven by a moving rotational trap (supplemental material). The total system dimensionality $d=2N$ was scaled from $d=10$ ($N=5$) up to $d=10,000$ ($N=5000$). We emphasize that this specific benchmark is designed  to measure computational throughput; it does not assess long-horizon physical accuracy at these dimensions. However, the physical fidelity of our framework in high-dimensional regimes has been independently verified up to $d=100$ ($N=50$) at $\lambda=0$ on the harmonically coupled system.

\onecolumngrid
\clearpage

\begin{center}
    \textbf{ \large Supplemental Material for: 
    Dynamical Partition Functions of Stochastic Dynamics via Variational Flows}\\[.3cm]
    Zequn Lin$^{1,2,*}$ and Ying Tang$^{3,4,\dagger}$ \\[.1cm]
    {\small \it 
    \parbox{0.9\textwidth}{\centering
    $^1$Department of Physics, Fudan University, Shanghai 200433, China\\
    $^2$Center for Interdisciplinary Studies, Westlake University, Hangzhou 310030, China\\
    $^3$Institute of Fundamental and Frontier Sciences,\\ University of Electronic Science and Technology of China, Chengdu 611731, China\\
    $^4$Non-classical Information Science Basic Discipline Research Center of Sichuan Province, \\University of Electronic Science and Technology of China, Chengdu 611731, China
    }}
\end{center}
\vspace{2em}

\setcounter{equation}{0}
\setcounter{figure}{0}
\setcounter{table}{0}
\setcounter{section}{0}
\setcounter{page}{1}

\makeatletter
\renewcommand{\theequation}{S\arabic{equation}}
\renewcommand{\thefigure}{S\arabic{figure}}
\renewcommand{\thetable}{S\arabic{table}}
\renewcommand{\thesection}{S\arabic{section}}
\renewcommand{\bibnumfmt}[1]{[S#1]}
\renewcommand{\citenumfont}[1]{S#1}
\makeatother

\section{Local Energy}

To derive the evolution equations (Eq.~(3)), we define the local energy as $E_{\text{loc}}(\mathbf{x}) \equiv p_\theta^{-1}\mathcal{L}_\lambda p_\theta$.  Under the Feynman-Kac formalism, the unnormalized density solving the tilted Fokker-Planck equation (FPE) is decomposed as $\rho(\mathbf{x}, t) = Z_t(\lambda) p_\theta(\mathbf{x}, t)$. Here, $Z_t(\lambda)$ is the scalar dynamical partition function and $p_\theta$ is the normalized probability density. Substituting this decomposition into the tilted FPE ($\partial_t \rho = \mathcal{L}_\lambda \rho$) and dividing by $\rho$ yields:

\begin{equation}\frac{\partial_t Z_t}{Z_t} + \frac{\partial_t p_\theta}{p_\theta} = \frac{\mathcal{L}_\lambda p_\theta}{p_\theta} \equiv E_{\text{loc}}(\mathbf{x}).\end{equation}

Taking the spatial expectation over $p_\theta$, and noting that $\int \partial_t p_\theta d\mathbf{x} = 0$, decouples the dynamics of the probability density from the global scaling factor:

\begin{equation}\partial_t \ln Z_t(\lambda) = \mathbb{E}_{\mathbf{x} \sim p_\theta} \left[ E_{\text{loc}}(\mathbf{x}) \right].\end{equation}

Physically, $E_{\text{loc}}(\mathbf{x})$ is the local growth rate of path weights at $\mathbf{x}$. Its expectation yields the generation rate of the dynamical partition function. Algorithmically, $E_{\text{loc}}(\mathbf{x})$ guides the natural gradient descent of the neural network parameters. Using the time-dependent variational principle (TDVP), we project the tilted dynamics onto the neural tangent space by minimizing the Kullback-Leibler divergence. The parameter evolution is:

\begin{equation}\dot{\theta} = \mathbf{F}_\theta^{-1} \mathbb{E}_{\mathbf{x} \sim p_\theta} \left[ (\nabla_\theta \ln p_\theta(\mathbf{x})) E_{\text{loc}}(\mathbf{x}) \right],\end{equation} where $\mathbf{F}_\theta$ is the Fisher information matrix (FIM). Phase space regions with high local energy therefore dominate the parameter updates, deforming the flow model to match rare event fluctuations. 
Using $E_{\text{loc}}(\mathbf{x})$ eliminates spatial integrals over a grid. 
We estimate it via independent samples from $\mathbf{x} \sim p_\theta$, as it requires only pointwise evaluations of the drift, the diffusion matrix, and the neural Laplacian.

\section{Numerical Implementation of the TDVP Optimization}

In the continuous-time variational formulation, the evolution of the normalizing flow parameters $\theta$ is governed by the natural gradient descent equation:
\begin{equation}
    \mathbf{F}_\theta \dot{\theta} = \mathbf{h}_\theta,
\end{equation}
where $\mathbf{h}_\theta = \mathbb{E}_{\mathbf{x} \sim p_\theta} \left[ (\nabla_\theta \ln p_\theta) E_{\text{loc}}(\mathbf{x}) \right]$ is the driving force vector, and $\mathbf{F}_\theta = \mathbb{E}_{\mathbf{x} \sim p_\theta} \left[ (\nabla_\theta \ln p_\theta) (\nabla_\theta \ln p_\theta)^\top \right]$ is the FIM. 

For a neural network with $N_{\text{param}}$ parameters, explicitly constructing and inverting the dense $N_{\text{param}} \times N_{\text{param}}$ FIM incurs prohibitive $\mathcal{O}(N_{\text{param}}^3)$ computational and $\mathcal{O}(N_{\text{param}}^2)$ memory costs. Following state-of-the-art optimization techniques in large-scale deep learning  and neural quantum states, we implement a matrix-free iterative solver using the conjugate gradient (CG) algorithm.

The CG method never requires the explicit instantiation of $\mathbf{F}_\theta$; it requires a function that computes the Fisher-vector product (FVP) $\mathbf{F}_\theta \mathbf{v}$ for an arbitrary parameter-space vector $\mathbf{v}$. In our implementation, we implicitly evaluate this inner product utilizing automatic differentiation or centered finite differences. By perturbing the network weights by a small scalar $\epsilon$, we approximate the directional derivative of the log-likelihood as $J_{\mathbf{v}}(\mathbf{x}) \approx (2\epsilon)^{-1} [ \ln p_{\theta + \epsilon \mathbf{v}}(\mathbf{x}) - \ln p_{\theta - \epsilon \mathbf{v}}(\mathbf{x}) ]$. Once $J_{\mathbf{v}}(\mathbf{x})$ is evaluated for a batch of samples, we center it via $\bar{J}_{\mathbf{v}} = J_{\mathbf{v}} - \mathbb{E}[J_{\mathbf{v}}]$ to reduce variance, and execute a standard backward pass to extract the  FVP:
\begin{equation}
    \mathbf{F}_\theta \mathbf{v} \approx \nabla_\theta \left( \mathbb{E}_{\mathbf{x} \sim p_\theta} \left[ \ln p_\theta(\mathbf{x}) \bar{J}_{\mathbf{v}}(\mathbf{x}) \right] \right).
\end{equation}
This matrix-free paradigm acts as an implicit Hessian-vector product mechanism, reducing the parameter-update complexity to  $\mathcal{O}(N_{\text{param}})$ per CG step.

Because the FIM is estimated using a finite Monte Carlo ensemble, it frequently possesses zero or near-zero eigenvalues, rendering the matrix ill-conditioned. To guarantee strict positive-definiteness and stabilize the natural gradient projection in flat regions of the probability manifold, we apply Tikhonov regularization. This is  achieved by introducing a small diagonal shift $\epsilon_{\text{shift}}$ to the FIM during the implicit FVP computation: $\mathbf{F}_\theta \mathbf{v} \to \mathbf{F}_\theta \mathbf{v} + \epsilon_{\text{shift}} \mathbf{v}$.

\section{Tilted Operators in the Main Text}

In the main text, Eq.~(3) introduces the tilted Fokker-Planck operator for a purely state-dependent path observable $\mathcal{O}(\mathbf{x})$. However, for generalized thermodynamic quantities such as topological particle currents and entropy production, the path observables explicitly depend on the stochastic increments $d\mathbf{x}_t$. To  formulate a unified continuous-time variational inference framework for all observables, we must derive their associated forward tilted Fokker-Planck equations using the Cameron-Martin-Girsanov transformation \cite{chetrite2015nonequilibrium}.

\subsection{A. General Derivation via Girsanov Transformation}

Consider the overdamped Langevin equation $d\mathbf{x}_t = \mathbf{f}(\mathbf{x}_t) dt + \sqrt{2\mathbf{D}} d\mathbf{W}_t$ and a generalized path observable defined via Stratonovich integration:
\begin{equation}
O_t = \int_0^t f(\mathbf{x}_\tau, \tau) d\tau + \int_0^t \mathbf{g}(\mathbf{x}_\tau, \tau) \circ d\mathbf{x}_\tau.
\end{equation}
According to the large deviation theory of Markov processes, the backward tilted generator $\mathcal{L}_{-\lambda}^{bwd}$ governing the SCGF with a tilting parameter $-\lambda$ is obtained via the Girsanov transformation as:
\begin{equation}
\mathcal{L}_{-\lambda}^{bwd} = \mathbf{f} \cdot (\nabla - \lambda\mathbf{g}) + (\nabla - \lambda\mathbf{g}) \cdot \mathbf{D} (\nabla - \lambda\mathbf{g}) - \lambda f.
\end{equation}
To evaluate the dynamical partition function $Z_t(\lambda) = \int \rho(\mathbf{x}, t) d\mathbf{x}$, we require the forward tilted Fokker-Planck operator $\mathcal{L}_\lambda$ that governs the unnormalized density $\partial_t \rho = \mathcal{L}_\lambda \rho$. This operator is precisely the formal adjoint of the backward generator, $\mathcal{L}_\lambda = (\mathcal{L}_{-\lambda}^{bwd})^\dagger$. Taking the adjoint yields the general operator form:
\begin{equation}
\label{eq:general_tilted_fpe}
\mathcal{L}_\lambda \rho = (-\nabla - \lambda\mathbf{g}) \cdot \mathbf{D} (-\nabla - \lambda\mathbf{g}) \rho - (\nabla + \lambda\mathbf{g}) \cdot (\mathbf{f}\rho) - \lambda f \rho.
\end{equation}
This universal formula generates the exact tilted FPEs for all three specific examples studied in the main text.

\subsection{B. Finite-Time Jarzynski Equality ($d=2$)}

For the 2D Ornstein-Uhlenbeck process, the target observable is the work $W_t$. In stochastic thermodynamics, the injected power done by the external protocol $\boldsymbol{\mu}(t)$ is defined as the partial time derivative of the effective potential $\Phi(\mathbf{x}, t) = \frac{1}{2}(\mathbf{x} - \mathbf{x}^*_t)^\top \mathbf{U} (\mathbf{x} - \mathbf{x}^*_t)$, where $\mathbf{x}^*_t = \mathbf{F}^{-1}\boldsymbol{\mu}(t)$. 

Evaluating the injected power $\dot{W}(\tau) = \frac{\partial \Phi}{\partial \tau}$, we define the state-dependent path observable:
\begin{equation}
\mathcal{O}(\mathbf{x}, t) = \mathbf{x}^\top \mathbf{U}^\top \mathbf{F}^{-1} \dot{\boldsymbol{\mu}}(t).
\end{equation}
Because thermodynamic work in this continuous-state system depends solely on the state $\mathbf{x}_\tau$ and time $\tau$, it contains no stochastic increments. Consequently, we have $\mathbf{g}(\mathbf{x}) = 0$ and $f(\mathbf{x}, t) = \mathcal{O}(\mathbf{x}, t)$. 

Setting $\mathbf{g} = 0$ in Eq.~(\ref{eq:general_tilted_fpe}) reduces the tilted operator to the standard Feynman-Kac form of Eq.~(3):
\begin{equation}
\mathcal{L}_\lambda \rho = \nabla \cdot (\mathbf{D}\nabla \rho) - \nabla \cdot (\mathbf{f}\rho) - \lambda \mathcal{O}(\mathbf{x}, t) \rho.
\end{equation}

\subsection{C. Topological Current on the Unit Circle ($d=1$)}

For the 1D periodic diffusion process, the dynamics are driven by $\mathbf{f}(x) = -V'(x) + \gamma$ with a scalar diffusion $D=1$. The target observable is the time-integrated topological current $O_t = \int_0^t 1\circ  dx_\tau$. 

Mapping this to our general formulation, we have $f(x) = 0$ and $\mathbf{g}(x) = 1$. Substituting these into Eq.~(\ref{eq:general_tilted_fpe}) yields the forward tilted operator:
\begin{equation}
\mathcal{L}_\lambda \rho = (-\partial_x - \lambda)^2 \rho - (\partial_x + \lambda)(\mathbf{f}\rho).
\end{equation}
Expanding the spatial derivatives explicitly, we recover the exact tilted FPE:
\begin{equation}
\mathcal{L}_\lambda \rho = \partial_x^2 \rho - \partial_x (\mathbf{f}\rho) + 2\lambda \partial_x \rho - \lambda \mathbf{f} \rho + \lambda^2 \rho.
\end{equation}

\subsection{D. High-Dimensional Entropy Production ($d=16$)}

The dynamics are governed by $\mathbf{f}(\mathbf{x}) = -\nabla V(\mathbf{x}) + \mathbf{b}(\mathbf{x})$ with diffusion $\mathbf{D} = \epsilon \mathbf{I}$. The cutoff function is defined as $\eta(r)=1$ for $r\le 1$, $\eta(r)=0$ for $r\ge 2$, and $\eta(r)=\frac{1}{2}\left[1+\cos\bigl(\pi(r-1)\bigr)\right]$ for $1<r<2$. For the observable $O_t = \frac{1}{\epsilon} \int_0^t \mathbf{b} \circ d\mathbf{x}_\tau$, the generator of the tilted  $\mathcal{L}_\lambda$ is defined by:
\begin{equation}
\mathcal{L}_\lambda \rho = \epsilon \left(-\nabla - \frac{\lambda}{\epsilon}\mathbf{b}\right)^2 \rho - \left(\nabla + \frac{\lambda}{\epsilon}\mathbf{b}\right) \cdot (\mathbf{f}\rho).
\end{equation}
Expanding all vector calculus operators, we obtain the tilted operator:
\begin{equation}
\mathcal{L}_\lambda \rho = \epsilon \nabla^2 \rho + \nabla V \cdot \nabla \rho + (2\lambda - 1) \mathbf{b} \cdot \nabla \rho + (\nabla^2 V)\rho + (\lambda - 1)(\nabla \cdot \mathbf{b})\rho + \frac{\lambda(\lambda - 1)}{\epsilon} |\mathbf{b}|^2 \rho + \frac{\lambda}{\epsilon} (\mathbf{b} \cdot \nabla V) \rho.
\label{eq:raw_operator}
\end{equation}

\section{Physical System Setup for the Scaling Benchmark}

In the end matter of the main text, we present a computational throughput test demonstrating $\mathcal{O}(d^2)$ polynomial scaling up to $d=10,000$. The specific nonequilibrium system used for this benchmark consists of $N$ globally coupled two-dimensional particles ($d=2N$). The stochastic dynamics for the $i$-th particle ($\mathbf{x}_t^{(i)} \in \mathbb{R}^2$) are governed by the following system of coupled stochastic differential equations:
\begin{equation}
    d\mathbf{x}_t^{(i)} = \left[ \boldsymbol{\beta}(t) - \left( 1 - \alpha + \frac{\alpha}{N} \right) \mathbf{\Gamma} \mathbf{x}_t^{(i)} - \frac{\alpha}{N} \sum_{j \neq i}^N \mathbf{x}_t^{(j)} \right] dt + \sqrt{2D} \, d\mathbf{W}_t^{(i)},
\end{equation}
where $i = 1, \dots, N$, and $\alpha \in (0,1)$ dictates the global coupling strength among the particles. To introduce a persistent probability current and explicitly break detailed balance, we apply an asymmetric rotational matrix $\mathbf{\Gamma}$ strictly to the self-interaction term:
\begin{equation}
    \mathbf{\Gamma} = \begin{pmatrix} 1 & -1 \\ 1 & 1 \end{pmatrix}.
\end{equation}
The external driving protocol $\boldsymbol{\beta}(t)$ is a periodic circular trap shared globally by all particles:
\begin{equation}
    \boldsymbol{\beta}(t) = a \begin{pmatrix} \cos(\pi \omega t) \\ \sin(\pi \omega t) \end{pmatrix}.
\end{equation}
This formulation constructs a dense,  non-symmetric global drift matrix $\mathbf{F} \in \mathbb{R}^{d \times d}$, forcing the neural ansatz to resolve a highly complex, non-gradient probability landscape. In our scaling tests, we utilize the specific parameters: $a=2.0$, $\omega=1.0$, $D=0.25$, and $\alpha=0.5$.

\section{FPE Tracking Accuracy Validation in High Dimensions ($d=100$)}

While the $d=10,000$ test in the main text demonstrates computational speed, we independently verify the physical accuracy of our framework using $N=50$ interacting particles ($d=100$). We choose a benchmark introduced by Boffi et al.\cite{Boffi_2023} that has an exact solution. We consider $N=50$ two-dimensional particles that experience a pairwise harmonic repulsion while being attracted to a common moving harmonic trap. This configuration gives rise to a 100-dimensional Fokker-Planck equation. The stochastic dynamics of the $i$-th physical particle are governed by the following coupled stochastic differential equations:
\begin{equation}
dX_t^{(i)} = (\beta_t - X_t^{(i)})dt + \alpha \left( X_t^{(i)} - \frac{1}{N}\sum_{j=1}^N X_t^{(j)} \right)dt + \sqrt{2D}dW_t^{(i)}, \quad i=1,\dots,N,
\end{equation}
where $\alpha \in (0,1)$ is a fixed coefficient dictating the magnitude of the inter-particle repulsion, and $D$ is the diffusion coefficient. The position of the moving trap follows a circular trajectory defined as $\beta_t = a(\cos (\pi \omega t), \sin (\pi \omega t))^\top$. Following the exact configuration in  \cite{Boffi_2023}, we set the system parameters to $a=2$, $\omega=1$, $D=0.25$, and $\alpha=0.5$. The particles are initialized from a Gaussian distribution centered at the initial trap position $\beta_0$ with variance $\sigma_0^2 = 0.25$.

Because it is a high-dimensional Ornstein-Uhlenbeck process with a Gaussian initial condition, the exact time-dependent probability density remains  Gaussian for all $t \ge 0$. Consequently, its evolution can be fully characterized by its mean and covariance matrix, both of which can be derived analytically. 

\begin{figure}[h!]
    \centering
    \includegraphics[width=\linewidth]{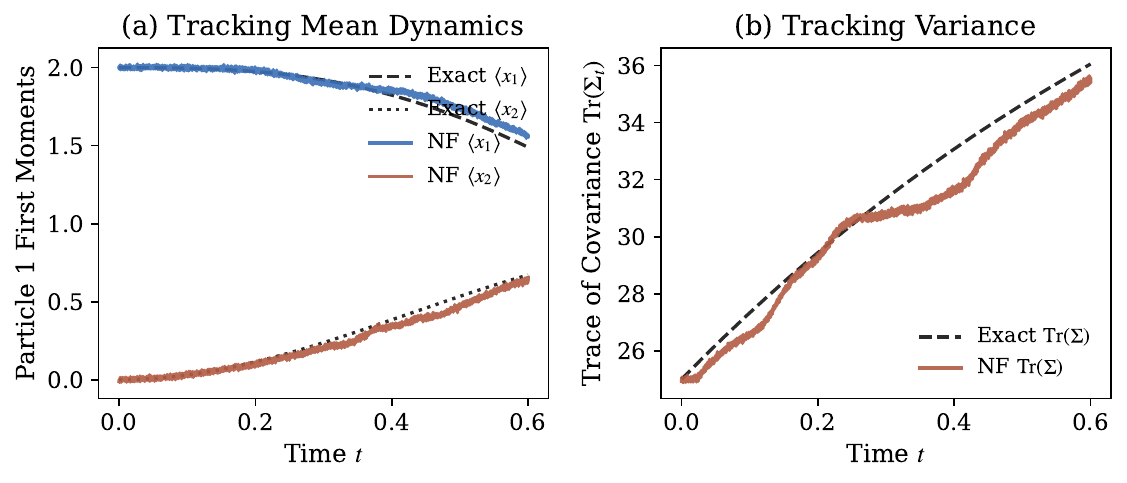}
    \caption{Validation of the neural variational framework on a 100-dimensional system. We simulate a system of $N=50$ harmonically interacting particles driven by a moving circular trap \cite{Boffi_2023}. (a) The time evolution of the first moments $\langle x_1 \rangle$ and $\langle x_2 \rangle$ for a representative particle. (b) The trace of the system's covariance matrix $\text{Tr}(\Sigma_t)$. In both panels, the solid colored lines denote the numerical estimates obtained directly from the normalizing flow via the TDVP, while the black dashed and dotted lines represent the exact solutions.}
    \label{fig:50N}
\end{figure}

As shown in FIG.~\ref{fig:50N}, the neural variational ansatz tracks the exact analytical transient moments and the trace of the covariance matrix $\text{Tr}(\Sigma_t)$ over long simulation horizons. This confirms that the mesh-free optimization remains stable in high-dimensional regimes.

\FloatBarrier  
\section{Normalizing Flow Architecture and Optimization Details}

Across all experiments, the neural networks parameterizing the affine and spline coupling layers are implemented as multi-layer perceptrons (MLPs). Each MLP consists of two hidden layers with the hyperbolic tangent (Tanh) activation function. The final linear layer of each MLP is initialized with near-zero weights to ensure that the initial normalizing flow closely approximates the identity transformation. 

\subsection{RealNVP}

To ensure the framework is self-contained and our numerical results are fully reproducible, we explicitly detail the mathematical formulation of the normalizing flow used in our Euclidean experiments (Benchmarks A, C, D, and E). We parameterize the probability density $p_\theta(\mathbf{x})$ using the RealNVP architecture \cite{dinh2016density}, which constructs a bijective mapping $\mathbf{x} = g_\theta(\mathbf{z})$ from a simple latent distribution.

The full transformation $g_\theta$ is composed of a sequence of invertible affine coupling layers. In each coupling layer, the input vector $\mathbf{v}$ is partitioned into two disjoint subsets using a binary mask $\mathbf{m} \in \{0, 1\}^d$: the unchanged part $\mathbf{v}^A = \mathbf{m} \odot \mathbf{v}$ and the transformed part $\mathbf{v}^B = (\mathbf{1} - \mathbf{m}) \odot \mathbf{v}$, where $\odot$ denotes the element-wise Hadamard product. The forward generative transformation to the output $\mathbf{u}$ is defined as:
\begin{align}
    \mathbf{u}^A &= \mathbf{v}^A, \\
    \mathbf{u}^B &= \mathbf{v}^B \odot \exp\left( \mathbf{s}_\theta(\mathbf{v}^A) \right) + \mathbf{t}_\theta(\mathbf{v}^A),
\end{align}
where $\mathbf{s}_\theta$ (scale) and $\mathbf{t}_\theta$ (translation) are arbitrary, potentially complex neural networks. To  respect the topological requirements and ensure all degrees of freedom interact, the binary mask $\mathbf{m}$ alternates between even and odd indices across successive coupling layers. The Jacobian determinant of this transformation is simply the product of the exponentiated scale terms, rendering the exact likelihood evaluation  efficient: $\log |\det \mathbf{J}| = \sum \mathbf{s}_\theta(\mathbf{v}^A)$. Here, both $\mathbf{s}_\theta$ and $\mathbf{t}_\theta$ are parameterized by two-layer MLPs with Tanh activations. 

\subsection{Wrapped Spline Flow}

For systems on  $\mathbb{S}^1$, standard flows violate boundary conditions. We lift the 1D physical state $x_{\text{phys}}$ into $\mathbb{R}^2$ via an auxiliary variable $x_{\text{ghost}}$, and apply rational-quadratic spline (RQS) coupling layers \cite{durkan2019neural} with $K=8$ bins on $[-5, 5]$, parameterized by 64-unit Tanh MLPs.

To enforce periodicity, the wrapped density sums over periodic images:
\begin{equation}
    p_{\text{wrap}}(x_{\text{phys}}, x_{\text{ghost}}) \approx \sum_k p_{\text{flow}}(x_{\text{phys}} + k, x_{\text{ghost}}).
\end{equation}
Sampling is performed by drawing from the spline flow and projecting $x_{\text{phys}} = x_{\text{flow}} \pmod 1$.

\subsection{A. Finite-Time Jarzynski Equality ($d=2$)}

For the 2D Ornstein-Uhlenbeck process, we employed a standard RealNVP architecture mapping from a standard normal base distribution $\mathcal{N}(0, \mathbf{I})$. The network comprises 2 affine coupling layers with a hidden dimension of 64. Prior to the physical evolution, we pre-trained the network for 500 steps using the Adam optimizer with a learning rate of $10^{-3}$ to match the initial Gaussian density. During the physical evolution, the system was simulated using a time step of $\delta t = 10^{-4}$ for a total of $N_{\text{step}} = 300$ steps, utilizing an ensemble of $200,000$ samples. The TDVP optimization was performed with a learning rate of $\eta = 10^{-4}$, a diagonal shift of $\epsilon_{\text{shift}} = 10^{-3}$, and a maximum of $N_{\text{CG}} = 20$.

\subsection{B. Topological Current on the Unit Circle ($d=1$)}

To satisfy the periodic boundary conditions on $\mathbb{S}^1$, we utilized a wrapped rational-quadratic spline flow consisting of 4 coupling layers and 8 bins, with a hidden dimension of 64. The system was integrated up to $t=0.5$ using a numerical time step of $\delta t = 2 \times 10^{-3}$ and an ensemble size of $20,000$. The TDVP projection was executed with a learning rate of $\eta = 1.0$ and up to $N_{\text{CG}} = 30$ iterations, alongside an annealed diagonal shift $\epsilon_{\text{shift}}$ to guarantee optimization stability.

\subsection{C. High-Dimensional Entropy Production ($d=16$)}

For the 16-dimensional system, we mapped the target density from a uniform base distribution defined on $[-0.15, 0.15]^{16}$ using 4 affine coupling layers with a hidden dimension of 64. The stochastic dynamics were simulated to $t=1.0$ with a time step of $\delta t = 10^{-3}$ and an ensemble consisting of $4096$ samples. The flow parameters were optimized using a learning rate of $\eta = 10^{-3}$ and a regularizing diagonal shift of $\epsilon_{\text{shift}} = 1 \times 10^{-3}$.

\subsection{D. High-Dimensional Tracking Benchmark ($d=100$)}

For the 100-dimensional system consisting of $N=50$ particles, we used a standard RealNVP architecture mapping from $\mathcal{N}(0, \mathbf{I})$. We employed a flow architecture consisting of 8 affine coupling layers with a hidden dimension of 32. The dynamics were integrated with $\delta t = 10^{-4}$ for $N_{\text{step}} = 6000$ steps using an ensemble of $1,000$ samples.  Convergence criteria were enforced for the conjugate gradient method, utilizing $\eta = 10^{-4}$, $\epsilon_{\text{shift}} = 10^{-4}$, and up to $N_{\text{CG}} = 50$.

\subsection{E. Scaling Analysis ($d=10$ to $10,000$)}

To  profile the computational complexity across varying dimensions, we maintained a standardized RealNVP architecture mapping from $\mathcal{N}(0, \mathbf{I})$. Since the primary objective here is to measure the single-step algorithmic wall-clock time rather than long-horizon physical accuracy, we utilized a shallower network comprising 4 affine coupling layers with a hidden dimension of 64. Measurements were conducted using a fixed ensemble size of $8,000$ samples. The TDVP projection was executed with a learning rate of $\eta = 10^{-3}$, a diagonal shift of $\epsilon_{\text{shift}} = 10^{-4}$, and a maximum of $N_{\text{CG}} = 30$ iterations.

\section{Exact Solution of the 2D Ornstein-Uhlenbeck Process}

To  validate our variational flow framework against nonequilibrium fluctuation theorems, we consider a two-dimensional Ornstein-Uhlenbeck process driven by a time-dependent protocol. The overdamped Langevin equation for the state vector $\mathbf{x}_t \in \mathbb{R}^2$ reads:
\begin{equation}
d\mathbf{x}_t = \left[ -\mathbf{F}\mathbf{x}_t + \boldsymbol{\mu}(t) \right] dt + \sqrt{2\mathbf{D}} \, d\mathbf{W}_t.
\end{equation}
We set the diffusion matrix to the identity $\mathbf{D} = \mathbf{I}$, and the constant drift matrix $\mathbf{F}$ is defined as:
\begin{equation}
\mathbf{F} = \begin{pmatrix} 1 & -1 \\ 1 & 1 \end{pmatrix}.
\end{equation}
The skew-symmetric off-diagonal elements in $\mathbf{F}$ explicitly break detailed balance, inducing a rotational probability current. The external driving protocol $\boldsymbol{\mu}(t) = (\mu_1(t), 0)^\top$ operates exclusively along the first coordinate with a driving speed $K=20.0$, following a schedule:
\begin{equation}
\mu_1(t) = 
\begin{cases} 
20.0 t, & t \le 0.025 \\
0.5, & t > 0.025 
\end{cases}.
\end{equation}

We first evaluate the steady-state covariance matrix $\mathbf{\Sigma}$ by solving the Lyapunov equation $\mathbf{F}\mathbf{\Sigma} + \mathbf{\Sigma}\mathbf{F}^\top = 2\mathbf{D}$. Decomposing the drift matrix as $\mathbf{F} = \mathbf{I} + \mathbf{J}$, where $\mathbf{J} = [\begin{smallmatrix} 0 & -1 \\ 1 & 0 \end{smallmatrix}]$ is  skew-symmetric, one can easily verify that the exact solution yields $\mathbf{\Sigma} = \mathbf{I}$. Next, we define the anti-symmetric matrix $\mathbf{R} = \mathbf{F}\mathbf{D} - \mathbf{D}\mathbf{F}^\top = [\begin{smallmatrix} 0 & -2 \\ 2 & 0 \end{smallmatrix}]$, which measures the detailed balance violation. The steady-state covariance correction $\mathbf{Q}$ is determined by $\mathbf{F}\mathbf{Q} + \mathbf{Q}\mathbf{F}^\top = \mathbf{R}$. Assuming an anti-symmetric ansatz $\mathbf{Q} = [\begin{smallmatrix} 0 & -q \\ q & 0 \end{smallmatrix}]$, direct substitution yields $2q = 2$, revealing $\mathbf{Q} = [\begin{smallmatrix} 0 & -1 \\ 1 & 0 \end{smallmatrix}]$. Consequently, the effective potential characterizing the nonequilibrium steady state is $\Phi(\mathbf{x}) = \frac{1}{2}\mathbf{x}^\top\mathbf{U}\mathbf{x}$, where the potential matrix is formulated as $\mathbf{U} = (\mathbf{D} + \mathbf{Q})^{-1}\mathbf{F}$. Remarkably, since $\mathbf{D} + \mathbf{Q} = \mathbf{F}$, the effective potential matrix collapses to the identity, $\mathbf{U} = \mathbf{I}$.

Under the external protocol, the instantaneous minimum of this effective potential shifts to $\mathbf{x}^*_t = \mathbf{F}^{-1}\boldsymbol{\mu}(t)$. Explicit matrix inversion yields the exact trajectory of the potential minimum:
\begin{equation}
\mathbf{x}^*_t = \frac{1}{2} \begin{pmatrix} 1 & 1 \\ -1 & 1 \end{pmatrix}\begin{pmatrix} \mu_1(t) \\ 0 \end{pmatrix} = \begin{pmatrix} \mu_1(t)/2 \\ -\mu_1(t)/2 \end{pmatrix}.
\end{equation}
The exact free energy difference, serving as the analytical baseline for validating the Jarzynski equality ($\langle e^{-W_t} \rangle = e^{-\Delta F}$), is thus derived as:
\begin{equation}
\Delta F(t) = \frac{1}{2} (\mathbf{x}^*_t)^\top \mathbf{U} \mathbf{x}^*_t = \frac{1}{4} \mu_1(t)^2.
\end{equation}

Finally, when evaluating the dynamical partition function $\langle e^{-W_t} \rangle$ via the Feynman-Kac formalism, the path ensemble is exponentially tilted by the injected power observable. This tilting adds a deterministic drift in the evolution of the biased ensemble's mean $\mathbf{m}(t)$. Specifically, the additional drift induced by reweighting is $\mathbf{v}_{\text{ext}}(t) = \mathbf{\Sigma} \mathbf{U}^\top \mathbf{F}^{-1} \dot{\boldsymbol{\mu}}(t)$. Substituting $\mathbf{\Sigma} = \mathbf{I}$ and $\mathbf{U} = \mathbf{I}$ simplifies this macroscopic correction to $\mathbf{v}_{\text{ext}}(t) = \mathbf{F}^{-1} \dot{\boldsymbol{\mu}}(t) = \dot{\mathbf{x}}^*_t$.

The exact dynamics of the biased mean is  governed by $\dot{\mathbf{m}}(t) = -\mathbf{F}\mathbf{m}(t) + \boldsymbol{\mu}(t) + \dot{\mathbf{x}}^*_t$. Recognizing that the protocol force inherently balances the restoring force at the minimum ($\boldsymbol{\mu}(t) = \mathbf{F}\mathbf{x}^*_t$), the governing equation becomes a relaxation process for the macroscopic lag $\mathbf{y}(t) = \mathbf{m}(t) - \mathbf{x}^*_t$:
\begin{equation}
\dot{\mathbf{y}}(t) = -\mathbf{F}\mathbf{y}(t).
\end{equation}
Given the unbiased initial condition $\mathbf{m}(0) = \mathbf{x}^*_0 = \mathbf{0}$, the lag remains  zero throughout the rapid driving phase ($\mathbf{y}(t) \equiv \mathbf{0}$). Thus, the mean tracks the instantaneous potential minimum:
\begin{equation}
\mathbf{m}(t) = \mathbf{x}^*_t = \begin{pmatrix} \mu_1(t)/2 \\ -\mu_1(t)/2 \end{pmatrix}.
\end{equation}

\end{document}